\documentclass[manuscript]{acmart}

\theoremstyle{definition}
\newtheorem{definition}{Definition}[section]

\AtBeginDocument{%
  \providecommand\BibTeX{{%
    \normalfont B\kern-0.5em{\scshape i\kern-0.25em b}\kern-0.8em\TeX}}}

\renewcommand\footnotetextcopyrightpermission[1]{} 
\begin{document}

\title{Algebraic Structure of Blockchains: A Group-Theoretical Primer}

\author{Dongfang Zhao}
\email{dzhao@unr.edu}
\affiliation{%
  \institution{University of Nevada, Reno}
}

\begin{abstract}
Although recent advances of blockchain systems, notably in the form of cryptocurrency, have drawn tremendous interests from both researchers and practitioners,
limited studies existed toward the theoretical foundation of blockchains.
This paper presents the first study on the algebraic structure of blockchains with an emphasis on the internal properties under algebraic groups.
We axiomatically construct a blockchain group and derive some interesting properties that can be potentially taken into the design space and parametric analysis of real-world blockchain systems.
\end{abstract}

\maketitle

\section{Introduction}

While recent advances of blockchain systems, notably in the form of cryptocurrency,
have drawn tremendous interests from both researchers and practitioners~\cite{Wang_nsdi19,jlind_sosp19},
limited studies existed toward the theoretical foundation of blockchains before in 2017 Herlihy~\cite{mherlihy_podc17}, for the first time, brought up the connection between blockchains and distributed computing.
Admittedly, the original blockchain paper authored by Nakamoto~\cite{bitcoin} lacked formal proofs,
and yet, Bitcoin, as the banding name of public blockchains,
has proven to be a stable production system.
As we have witnessed in many other physical sciences, 
not every discipline started with the rigorous theory before the applications spread out vividly;
in fact, it often happened in a reverse way:
the internal law was found \textit{after} people observed many instances for a while.
We believe this would be the orbit of the development of blockchains:
we have witnessed many successful instances of blockchains,
now it might be the time for developing, or extracting, the internal laws of blockchains.

This paper strives to take a rigorous mathematical methodology to better understand blockchains' fundamental properties.
More specifically, we are interested in unraveling blockchain's algebraic structures---the very underneath commonality among production blockchain systems.
We are aware of a parallel study on blockchain theory from a \textit{game theory} standpoint (e.g., \cite{itsabary_ccs18}),
which focused on the \textit{rationality} and \textit{equilibrium} of blockchain nodes.
In contrast, our line of works aims to reveal the intrinsic laws of blockchain \textit{states} and \textit{actions} through algebraic structures. 

In the remainder of this paper,
we will provide a very high-level introduction to blockchains and algebraic groups.
Then, we will define some components that are needed for the axiomatic construction of blockchain groups.
After that, we will show that the construction indeed forms a well-defined algebraic group and more importantly,
derive some interesting properties that can be potentially considered in real-world blockchain design and analysis.
We finally conclude this paper and discuss future research directions.

\section{Background}

\subsection{Blockchkains}
A blockchain is a replicated database deployed to a distributed system.
By default, each node of the distributed system holds a full copy of the data, usually in a \textit{transactional} form.
Each replica of the database is organized as a \textit{hashed-linkedlist} of blocks of those transactional data such that the data cannot be compromised unless the entire chain is reconstructed from scratch,
which is computationally and financially prohibitive.
Since the overall system is \textit{replicated},
the system employs some \textit{consensus protocols} for all replicas to agree on.

A rough categorization of blockchains is based on its user membership:
if the system is publicly open to everyone, 
then the blockchain is called \textit{permissionless};
otherwise, the blockchain is called \textit{permissioned}.
Most of production blockchain systems are permissionless, 
and they are the emphasis of this paper.
At the writing of this paper, the most popular blockchain system Bitcoin, is comprised of about 9,800 nodes~\cite{bitcoin_scale}.
Each of such nodes might be in the form of a coalition of many physical machines,
namely a \textit{pool}~\cite{ieyal_sp15}.

\subsection{Algebraic Groups}

The algebraic \textit{group} is one of the most fundamental abstract algebraic structures---the basis of many more derived structures such as \textit{rings} and \textit{fields}.
Essentially, a group is a set of elements along with a binary operation defined on these elements:
The operation is \textit{associative} and there exists an \textit{inverse} for every element such that the product between an element and its inverse is an \textit{identify} element.
A group is usually constructed axiomatically:
first, define the set and binary operation and then show that all aforementioned properties are satisfied.

Of of the most interesting groups are called \textit{symmetric group},
denoted as $S_\Omega$,
whose elements are \textit{permutations} of the elements in set $\Omega$, and the binary operation is simply the function composition between two permutations.
Historically, early group-theoretical studies focused on the internal structure of $S_\Omega$ because it has a rich set of properties that can be applied to real-world disciplines such as computational chemistry and theoretical physics.
We will also leverage $S_\Omega$ in our study on blockchain groups.

\section{Definitions}

We use $N$ to denote the set of all nodes in the blockchain system.
We denote the set of all $r$ mining pools or clusters by $C = \{c_0, c_1, \dots, c_{r-1}\}$.
Of note, $c_0$ represents a dummy set of all the singleton nodes, 
i.e., those nodes that decide to mine the block individually without joining any pools.
In practice, we can assume $r \ll |N|$ as most nodes choose to join a pool.

\begin{definition}[Node-switch map $\varphi$]
We assume that any node can freely\footnote{Which is true for permissionless blockchains.} switch from one pool to another through a \textit{node-switch} map of an arbitrary node $n \in N$ from $C$ to $C$, $\varphi_n: C \rightarrow C$, such that:
\[
\varphi_n(c_i) = c_j \texttt{ if } n \in c_i \texttt{ and } n \texttt{ switches to pool } c_j, \texttt{ where } 0 \le i,j < r 
.\]
\end{definition}

Note that we do not require $i \not= j$ for a node-switch;
the map $\varphi$ is well defined even if the node stays in the same pool.
We will simply say $\varphi_n(i) = j$ if it is clear from the context that both $i$ and $j$ are indices for pools.
We use $\Phi_i$ to denote a set of node-switches covering all the nodes exactly once,
where $0 \leq i < |N|^r$.
Obviously, we have $|\Phi_i| = |N|$ for all $i$'s.
We call $\Phi_i$ a \textit{node-switch set} (of index $i$).

\begin{definition}[Pool-update map $\sigma$]\label{def:pool_update}
When the membership of a specific pool is updated,
e.g., a new node joins or an existing node leaves,
we denote such an change as a \textit{pool-update},
which is, formally, defined as a map $\sigma: \Phi \rightarrow \Phi$.
\end{definition}

We define the set of all possible maps $\sigma$'s (among $\Phi$'s) as $\Omega$.
It should be clear that it is the map $\sigma$ between node switch sets $\Phi$'s, \textit{not} $\Phi$ \textit{per se}, serving as an element in $\Omega$.
As an analogy, in the well-known symmetric group $S_n$,
it is the permutation between series of numbers, rather than the series itself, being considered as the element of $S_n$.
We then define $\star$ as a function composition between two $\sigma$'s among $\Phi$'s.
Obviously, we have $|\Omega| = |N|^r$.
Formally:
\[
\sigma_1 \star \sigma_2 = \sigma_3 \texttt{ if for any } n \in N, 
\{ \varphi_n(i) = j \} \subseteq \sigma_1 
\wedge \{ \varphi_n(j) = k \} \subseteq \sigma_2 
\Rightarrow \{ \varphi_n(i) = k \} \subseteq \sigma_3
.\]
Note that $\star$ is well defined because for any $\sigma$, all elements in $N$ appear on some source pools (cf. Def.~\ref{def:pool_update}, $|\Phi| = |N|$).
If the context is clear, we will simply write $\sigma_1 \sigma_2$ to denote the two elements in $\Omega$ operate under $\star$.

Now we are ready to show that set $\Omega$ forms an algebraic group under $\star$.

\section{Group-Theoretical Internals of Blockchains}

\subsection{Axiomatic Construction}

It boils down to demonstrating the following axioms for $\Omega$ to be a group under $\star$:
\begin{itemize}
    \item Operation $\star$ is associative: for any $\sigma_i \in \Omega$, $\sigma_j \in \Omega$, and $\sigma_k \in \Omega$, we have $\sigma_i \star (\sigma_j \star \sigma_k) = (\sigma_i \star \sigma_j) \star \sigma_k$;
    
    \item An identity element, denoted as $e \in \Omega$, exists such that for any $\sigma \in \Omega$, we have $e \star \sigma = \sigma \star e = \sigma$;
    
    \item For any $\sigma \in \Omega$, there exists an inverse counterpart, denoted as $\sigma^{-1}$, such that $\sigma \star 
    \sigma^{-1} = \
    \sigma^{-1} \star \sigma = e$.
\end{itemize}
We will show that all aforementioned axioms hold.

\subsubsection{Associativity}

Let $n \in N$ be any node in the blockchain. 
Let $\varphi_n(a) = b$ be any node switch in map $\sigma_k$.
Recall that there are a total of $|N|$ node switches in $\sigma_j$.
By definition of $\sigma$, there must exist \textit{one and only one} node switch from $b$ in $\sigma_k$: $\varphi_n(b) = c$, where $c$ is the destination pool $C_c$.
Then by definition of function association, we know $\varphi_n(a) = c$ is in $\sigma_j \star \sigma_k$.
Now, without loss of generality, let $\varphi_n(d) = a$ be a node switch in $\sigma_i$.
By definition of $\star$, we have that $\varphi_n(d) = c$ is a node switch in $\sigma_i \star (\sigma_j \star \sigma_k)$.
Note that both $d$ and $a$ are arbitrary indices of pools.

Similarly, if we know $\varphi_n(d) = a$ is a node switch in $\sigma_i$ and $\varphi_n(a) = b$ be a node switch in $\sigma_j$, respectively,
we then know that $\varphi_n(d) = b$ be a node switch in $\sigma_i \star \sigma_j$.
Consequently, if we know $\varphi_n(b) = c$ is a node switch in $\sigma_k$,
then we have, again, $\varphi_n(d) = c$ is a node switch in $\sigma_i \star (\sigma_j \star \sigma_k)$.

We thus have shown that $\star$ is associative in $\Omega$.

\subsubsection{Identity}

We construct $e$ as a $\sigma$ with its $|N|$ elements each of which is a trivial node switch: $\varphi_n(i) = i$ for all $n \in N$.
Obviously, any pool update would be mapped to its original structure after applying $e$, regardless of both left- and right-side function composition.
It should be noted, again, that each element of $\Omega$ is the map over the set of $|N|$ node switches, not the node switches themselves.

\subsubsection{Inverse}

For an arbitrary $\sigma$, each of the $|N|$ node switches can be written as $\varphi_n(i) = j$.
Because $\Omega$ is the set including all the possible maps between pool-updates,
there must exist a unique $\sigma^{-1}$ whose elements can be exactly written in the this form: $\varphi_n(j) = i$.
Then, for $\sigma \star \sigma^{-1}$, each node switch follows $\varphi_n(i) = i$, comprising $e$;
similarly, for $\sigma^{-1} \star \sigma$, each node switch follows $\varphi_n(j) = j$, again, comprising $e$.

\textbf{Remark.}
By construction, the group constructed from $\Omega$ and $\star$ is not \textit{commutative}, or \textit{non-abelian} in the literature of group theory.
We will denote such a blockchain group as $B_{n, r}$, where $n = |N|$ and $r = |C|$.
From the above discussion, we know the order of $B_{n,r}$ is $|B_{n,r}| = n^r$.


\subsection{Algebraic Properties and Applications}

This section presents some important properties implied by the non-abelian blockchain group $B_{n,r}$.

\subsubsection{Subgroups, lattices, normal subgroups, and kernels}

One of the most notable properties exhibited by $B_{n,r}$ lies at its order $n^r$.
We thus can rewrite it as follows:
\[
|B_{n,r}| = n^r = p_1^{\alpha_1 r} \times \dots \times p_k^{\alpha_k r}
\]
where $n = p_1^{\alpha_1} \dots p_k^{\alpha_k}$ where $p_i$ is a prime, $p_i < p_j$ for $1 \leq i < j \leq k$ and for any $1 \leq j \leq k$ we have $\alpha_j \geq 1$.
Note that, by this factorization, we have $p_i \nmid p_j^{\alpha_j r}$ for $i \not= j$ and $1 \leq i,j \leq k$,
where $a \nmid b$ reads $a$ cannot divide $b$.
It follows that 
\[
|B_{n,r}| = p_i^{\alpha_i r} \cdot m \texttt{ and } p_i \nmid m, \texttt{ where } 
m = |B_{n,r}| \cdot p_i^{-\alpha_i r}, 1 \leq i \leq k
,\]
which is exactly the form well studied by Sylow's Theorem:
if a group can be written in this form,
we know that there must exist a subgroup of order $p_i^{\alpha_i r}$ in $B_{n,r}$, where $1 \leq i \leq k$.
Since there $k$ such primes,
we know that $B_{n,r}$ has at least $k$ subgroups.
Consequently, we know that $B_{n,r}$ has a nontrivial lattice of subgroups.
This result itself could be useful for applications such as cryptography,
potentially leading to a new interdisciplinary research area:
leveraging blockchain's internal algebraic structure for encryption.

It would be highly useful if we could know how many of these subgroups are \textit{normal},
each of which essentially corresponds to a \textit{kernel} of $B_{n,r}$ that is widely used in group-theoretical applications.
However, without instantiating of $n$ and $r$, it is not analytically feasible to give the solution,
and this is particularly true if $n$ or $r$ or both is a medium or large number.
Nonetheless, we want to point out that for small-order subgroups, 
one can leverage Cayley's Theorem and Corollary:
if $p$ is the smallest prime that divides $|B|$,
then for any subgroup $H$ of $B$, denoted as $H \leq B$, if $\frac{\displaystyle |B|}{\displaystyle |H|} = p$,
then $H$ is normal, denoted as $H \trianglelefteq B$.
Therefore, for small $p$, such as 2 that is very likely included in the prime series of factors $(p_1, p_2, \dots, p_k)$ of $|B_{n,r}|$,
we can determine whether a subgroup $H$ by checking $n^r = 2|H|$.
If so, we will then have a lot of important applications built upon the kernel $H$.

\subsubsection{Coset order and element order}

According to Lagrange's Theorem,
the number of \textit{cosets} of a subgroup $H$ in $B_{n,r}$,
essentially the number of possible translates of $H$ from any $\sigma \in \Omega$,
can be calculated as $\frac{\displaystyle n^r}{\displaystyle |H|}$.
This can be translated into the blockchain network as:
if we know an active subset of $\sigma$'s included in a subgroup $H \leq G$,
then we can quickly determine exactly the number of (much fewer) possibilities that the node switches can lead to.

Next, we show that some elements (i.e., $\sigma$'s) have interesting \textit{cyclic} properties.
This is particularly useful by noting that $B$ is not cyclic in general.
According to Cauchy's Theorem:
if prime number $p$ divides $B$, then $B$ must have an element of order $p$.
Essentially, this means that there are at least $k$ elements in $B$, such that each of these elements that would degenerate to the identity element $e$.
More formally, we have 
\[
\sigma_i^{p_i} = \underbrace{\sigma_i \star \dots \star \sigma_i}_{p_i} = e, 1 \leq i \leq k
.\]
Intuitively, this means that some multiplications of pool-update maps would eventually result in the trivial map---no node switch at all.

\subsubsection{Homomorphism to Symmetric Group $S_\Omega$}

We conclude this section with a sketch about the intrinsic relationship between $B_{n,r}$, and the well-studied group $S_n$.
As a starting point, we want to reemphasize Cayley's Theorem that states:
every group $G$ is \textit{isomorphic} to a subgroup of a symmetric group $S_|G|$.
That is to say, $B_{n,r}$ is structurally identical, up to the operation $\star$ and a one-to-one mapping, 
to a subgroup of the well-understood symmetric group $S_{B_{n,r}}$.
More specifically, we know that $B_{n,r}$ is part of a lattice of the symmetric group $S_B$.
Historically, a subgroup of a symmetric group is also called a \textit{permutation group}.
However, it should be noted that working directly on a $S_B$ is prohibitively costly:
the order of $S_B$ is $|S_{B_{n,r}}| = (n^r)!$.
Using Stirling's approximation, we have:
\[
|S_{B_{n,r}}| = (n^r)! \sim \sqrt{2\pi n^r} \cdot (\frac{n^r}{e})^{n^r}
.\]
If we have a 10-node tiny blockchain with two pools,
$n = 10$ and $r = 2$,
the order of the blockchain group is manageable: $|B_{10,2}| = 100$;
and yet, the corresponding symmetric group has a order of:
$|S_{B_{10,2}}| \approx \sqrt{200\pi} \cdot (\frac{100}{e})^{100}  > 2^{512}$,
which is computationally infeasible.
As a side note, the state-of-the-art hash function for many production blockchain systems SHA256 takes 512-bit inputs and return 256-bit outputs;
the order of a tiny-scale $B$ group thus, as we just showed, already hit such a high security level.
Therefore, in the following, we provide analytical insights on the relationship between $B_{n,r}$ and $S_n$.


The key correlation between $B_{n,c}$ and $S_n$ lies at the structure of \textit{pools} in the blockchain.
Although we differentiate the elements in the set of pools $C$ into $\{c_0, c_1, \dots, c_{r-1}\}$,
what really makes these elements different is its membership of $\sigma$'s between node-switch sets $\Phi$'s.
Specifically,
if a map $\sigma$ only updates the pool index with the node membership unchanged within each pool,
then the new blockchain is essentially a permutation of the original one up to the pool topology.
Formally, in any $\sigma: \Phi_i \mapsto \Phi_j$,
if for any subset $M \subseteq N$ and any $m \in M$ we have both $\{\varphi_m(a) = b\} \subseteq \Phi_i$ and $\{\varphi_m(a) = c\} \subseteq \Phi_j$.
With this constraint, the blockchain group $B$ degenerates to a symmetric group at the granularity of pools: $B_{n,r} \cong S_n$.
It should be noted that this result is not applicable to a general blockchain group.







\section{Final Remark}

This paper presents the first study on the algebraic structure of blockchains with an emphasis on the internal properties under algebraic groups.
We axiomatically construct a blockchain group and derive some interesting properties that can be potentially taken into the design space and parametric analysis of real-world blockchain systems.
Specifically, 
we show that (i) a blockchain group, $B$,
comprises non-trivial subgroups and lattices that can be possibly leveraged for cryptography;
(ii) although $B$ is non-cyclic in general,
there exist cyclic elements in $B$,
which can help us reduce space in some scenarios; and
(iii) $B$ is homomorphic to the well-studied symmetric group $S_\Omega$ if some constraints hold,
thus opening the door to applying the wisdom of $S_\Omega$ to $B$.

Our future work lies in the development of algebraic structures among multiple blockchains.
For instance, the cross-blockchain transactions~\cite{dzhao_cidr20} might be analogous to group actions,
which can be possibly modeled by the \textit{orbits} with the \textit{conjugate} entities from distinct blockchains.
As another example, it would be worthwhile to explore the consequence of the \textit{primality} of the number of nodes,
where we might apply some number-theoretical techniques.

\section*{Acknowledgement}

This work is in part supported by the U.S. Department of Energy under contract number DE-SC0020455.
This work is also supported by a Google Cloud award and an Amazon research award.

\bibliographystyle{ACM-Reference-Format}
\bibliography{acmart}

\end{document}